\documentstyle[12pt]{article}
\begin{document}

\title{Four-terminal SQUID:Magnetic Flux Switching in Bistable State and Noise}
\author{R.de Bruyn \ Ouboter~${}^{a}$  , A.N.\ Omelyanchouk~${}^{b}$ 
\\ 
{\small {\em ${}^{a}$ Kamerlingh Onnes Laboratory,}}
\\
{\small {\em Leiden Institute of Physics,}}
\\
{\small {\em Leiden University,}}
\\
{\small {\em P.O. 9506, 2300 RA Leiden, The Netherlands}} 
\\
{\small {\em ${}^{b}$ B.Verkin Institute for Low Temperature Physics and
Engineering,}}
 \\
{\small {\em National Academy of Sciences of Ukraine,}}
\\
{\small {\em 47 Lenin Ave., 310164 Kharkov, Ukraine}}}
\date{}
\maketitle

\begin{abstract}
The effect of thermal fluctuations on the behaviour of a
4-terminal SQUID is investigated. The studied system consists of the 
Josephson 4-terminal junction with two terminals short-circuited by a 
superconducting ring and the other two form the transport circuit. The behaviour
of a 4-terminal SQUID is controlled by external parameters, the applied magnetic 
flux and the transport current. They determine the voltage in the transport
channel and the magnetic flux embraced in the ring. Within the numerical
model the noise-rounded current-voltage characteristics and the time dependence  of
the magnetic flux have been calculated. In some region of the control parameters
the 4-terminal SQUID is in the bistable state with two magnetic flux values.
The switching of the flux in bistable state produced by the thermal noise or
by the transport current is studied.

\end{abstract}
\newpage
\section{Introduction}
Presently the properties of superconducting microstructures attract
theoretical and experimental interest. Mostly it is due to the high level of
technology which permits the fabrication of the ultrasmall Josephson weakly
coupled systems with well defined geometry. Thus, the well known effects of
weak superconductivity \cite{jos,bar} now can be studied in the
situation approaching to the idealized theoretical models. From another
side, new physics (e.g. unconventional proximity effect,
reduced shot noise, Andreev's interferometry) appears
in small mesoscopic structures \cite{im}. The conventional Josephson effect in
microconstrictions also can be considered as mesoscopic phenomena \cite{bag}
produced by the macroscopic interference in the constriction of the
initially coherent Cooper pairs wave functions of the two massive
superconducting banks. The old idea\cite{lik,kok} of more complicated
Josephson weak link structures, the so called Josephson multiterminals \cite{vol1}, was
realized experimentally in ref.\cite{exp,fort} . In such structures the quantum interference takes
place in the intersection of several microbridges connecting superconducting
terminals. The specific multichannel interference effects were studied
theoretically and experimentally in the novel superconducting device, the
4-terminal SQUID controlled by the transport current \cite{oov1}. In this
system we have the interplay between the coherent current states in the
superconducting ring placed in the magnetic field and the Josephson effect
in the transport circuit. The external transport current $I$ and the applied
magnetic flux $\Phi _e$ are the control parameters which govern the behavior
of the 4-terminal SQUID. These two independent parameters determine the
response of the system: the total magnetic flux $\Phi $ in the ring and the voltage 
$V$ over the transport channel. It was shown \cite{oov1} that in the steady state
domain (when the transport current is less than the critical value $I_c(\Phi
_e)$) the nonlinear parametric coupling leads to a bistable state of the SQUID 
$\Phi (\Phi _e,I)$ for all values of the self-inductance of the ring.
Outside the steady state domain ($I>I_c)$ the resulting flux $\Phi $ becomes
an oscillating function of time whose amplitude, frequency and shape can be
smoothly controlled by the external parameters \cite{vol2,oov2}.

In the papers \cite{oov1,vol2,oov2} the theory of a 4-terminal SQUID was developed
without taking into account thermal fluctuations. The noise in the
system obviously plays an essential role. In particular it leads to the
magnetic flux switching in the bistable state, as was directly observed in
experiment \cite{bert}.

In the present paper we study the effects of thermal fluctuations on the
behavior of a 4-terminal SQUID. In Section 2, the basic equations used to
describe the system are presented. The thermal noise is considered by adding
the fluctuating voltages to the dynamical equations for the phases of the
superconducting order parameter. These equations were solved numerically
following the procedure described in \cite{tes}. In Section 3 the
calculations of the noise-affected current-voltage characteristics for
several values of applied magnetic flux are presented. The time dependence
of the magnetic flux in the ring stimulated by noise is studied in Section 4.

\section{Equations for the 4-terminal SQUID}

The 4-terminal SQUID which is controlled by the transport current is shown in Fig.1.
It consists of the Josephson 4-terminal junction with two terminals (3 and 4)
short-circuited by a superconducting ring and the other two (1 and 2) form
the transport circuit.

The set of equations describing the behavior of a 4-terminal SQUID in
an applied magnetic flux and a transport current was obtained in ref.\cite{oov1}. The
Josephson supercurrent in the $j$th filament (connecting the centre $o$ with the 
$j$th bank) expressed in terms of the phases of the superconducting order
parameter (related to the phase in the center $o$) has the form

\begin{equation}
I_j^s=\frac{\pi \Delta _0^2(T)}{4e{\it k} T_c}\frac
1{\sum\limits_{k=1}^41/R_k}\sum\limits_{k=1}^4\frac 1{R_jR_k}\sin (\varphi
_j-\varphi _k) 
\end{equation}

Here $\Delta _0$ is the gap in the bulk banks, $R_j$ is the normal
resistance of the $j$th filament. Let $V_j$ be the voltage in $j$th
terminal, related to the voltage in the centre ( $\sum\limits_jV_j/R_j=0$).
In the frame of the heavily damped resistively shunted model\cite{bar} we add to
supercurrents (1 ) the normal currents $I_j^n=V_j/R_j$ with $V_j=\frac \hbar
{2e}\stackrel{.}{\varphi }_j$. Thus for the total current flowing in $j$th
branch we have the expression in terms of the phases $\varphi _j$

\begin{equation}
\begin{array}{c}
I_j= 
\frac{\pi \Delta _0^2(T)}{4e{\it k} T_c}\frac
1{\sum\limits_{k=1}^41/R_k}\sum\limits_{k=1}^4\frac 1{R_jR_k}\sin (\varphi
_j-\varphi _k)+\frac \hbar {2eR_j}\frac{d\varphi _j}{dt}+\frac{\delta V_j(t)%
}{R_j}, \\ j=1,2,3,4. 
\end{array}
\end{equation}

The positive sign of $I_{j}$ (2) corresponds to the direction of the
current from the centre to $j$th bank. Note that in the resistively shunted
model the superconducting and normal currents are conserved separately and
equations (2 ) satisfy to conservation of the total current, $%
\sum\limits_{j=1}^4I_j=0$ .

The third term in the r.h.s. of equation ( 2) represents the thermal Johnson
noise of the normal resistance $R_j$. We assume that the time dependent
voltage noise sources are uncorrelated, each having a white voltage spectral
density $S_V=4{\it k} TR_j$.

The values of the currents $I_j$ are determined by the circuit implication of
the Josephson multiterminal. In the 4-terminal SQUID
configuration considered here (see Fig.1) they are connected to the external controlling
parameters, the given transport current $I$ and the applied magnetic flux $\Phi
_e$, in such a way that

\begin{equation}
I_1=-I,I_2=I,I_3=-J,I_4=J,J=(\Phi _e-\Phi
)/L, 
\end{equation}
where $J$ is the circulating current in the ring with self-inductance $L$ on
which a magnetic flux $\Phi _e$ is applied; The total embraced flux $\Phi $
is related to the phase difference between the terminals 4 and 3

\begin{equation}
\Phi =\frac \hbar {2e}(\varphi _4-\varphi _3). 
\end{equation}

Four equations (2 ) with the relations (3 ) and (4 ) constitute the coupled
system of differential equations for the phases $\varphi _j$ and describe the
dynamical behavior of the 4-terminal SQUID in the presence of thermal noise.
They have the integral of motion $\sum\limits_j\dot \varphi _j/R_j=C$ ,
where the value of the constant without loss of generality can be put to
zero. Thus we have three independent phases and it is convenient to
introduce new variables

\begin{equation}
\varphi _2-\varphi _1=\theta ,\varphi _4-\varphi _3=\phi ,\frac 12(\varphi
_1+\varphi _2)-\frac 12(\varphi _3+\varphi _4)=\chi . 
\end{equation}

In the following for simplicity we will consider the symmetric case $%
R_1=R_2=R_3=R_4=R/2$. We use the following dimensionless units: current in
units of $I_0=\pi \Delta _0^2/(4e{\it k} T_cR)$, voltage in units of $I_0R$
and time in units of $\hbar /(2eRI_0)$. The dynamical equations for the variables $%
\theta ,\phi ,\chi $ take the form

\begin{equation}
\begin{array}{c}
\frac{d\theta }{dt}=I-\frac 12\sin \theta -\sin \frac \theta 2\cos \frac
\phi 2\cos \chi +\delta v_\theta (t), \\ \frac{d\phi }{dt}=\frac{\phi
_e-\phi }{_{{\cal L}}}-\frac 12\sin \phi -\sin \frac \phi 2\cos \frac \theta
2\cos \chi +\delta v_\phi (t), \\ \frac{d\chi }{dt}=-\sin \chi \cos \frac
\phi 2\cos \frac \theta 2+\delta v_\chi (t), 
\end{array}
\end{equation}
where $\phi _e=\frac{2e}\hbar \Phi _e$ and ${\cal L}=\frac{2e}\hbar I_0L$ is the
dimensionless self-inductance. The ''random forces'' $\delta v_\theta
,\delta v_\phi $ , and $\delta v_\chi $ are related to the initial
dimensionless fluctuating voltages in eq.(2 ) :

\begin{equation}
\delta v_\theta =\delta v_1-\delta v_2,\delta v_\phi =\delta v_3-\delta
v_4,\delta v_\chi =-\frac 12(\delta v_1+\delta v_2)+\frac 12(\delta
v_3+\delta v_4). 
\end{equation}

The corresponding spectral densities in dimensionless units are

\begin{equation}
S_{v_\theta }=S_{v_\phi }=4\Gamma ,S_{v_\chi }=2\Gamma ,\Gamma =\frac{%
2e{\it k} T}{\hbar I_0}. 
\end{equation}

The system of the coupled equations (6) determines the time dependencies of the
phases $\theta (t),\phi (t)$ and $\chi (t)$ in the presence of thermal noise
at given values of the control parameters - the transport current $I$ and
the external magnetic flux $\phi _e$ , which are assumed to be independent of time.
The time derivatives of the phase differences determine the voltages between
different terminals:

\begin{equation}
V_{21}=\dot \theta ,V_{43}=\dot \phi ,\frac 12(V_{13}+V_{24})=\dot \chi . 
\end{equation}

\section{ Current-voltage characteristics in the presence of noise.}

We solve the equations (6) numerically by integrating the phases stepwise in
time with a step $\Delta t$. The Johnson noise voltages $\delta v(t)$ are
modelled by three uncorrelated trains of voltage pulses of constant duration 
$\Delta t$ and amplitudes $\delta v$. The amplitudes $\delta v$ are
Gaussian-distributed pseudo-random numbers of zero mean value with $<$ $\delta
v_\theta ^2>=<\delta v_\phi ^2>=2\Gamma /\Delta t$, $<$ $\delta v_\chi
^2>=\Gamma /\Delta t$ in accordance with the spectral densities (8). The value
of $\Gamma $ (8) characterizes the noise level in our system.

In this section we use the above formulated numerical model to calculate the
noise-affected current-voltage characteristics of the 4-terminal SQUID. The
observable quantity $V$, the voltage in the transport channel averaged over
time, was obtained from the solution of eq.(6) for the phase $\theta (t)$ : $%
V\equiv<~\frac{d\theta }{dt}~>_t$ .

To check the numerical techniques we first calculated numerically the
noise-rounded $IV$ characteristics for the single resistively shunted
Josephson junction. In this well known case the phase $\theta $ satisfy $%
\dot \theta =I-\sin \theta +\delta v_\theta (t)$ and from the corresponding
Fokker-Planck equation the average voltage across the junction was obtained
by Ambegaokar and Halperin \cite{amb} analytically. In Fig.2 we compare our
numerical results with the Fokker-Planck calculation, they are in good
agreement.

Noise-rounded $IV$ characteristics calculated from the full set of equations
(6) for two values of applied magnetic flux $\phi _e=0,\pi $ are
displayed in Fig.3 a,b. We choose ${\it L}$=1.7, $\Gamma =0.2$, close to
the values estimated in experiments \cite{bert}. Figure 4 illustrates the
behavior of current-voltage characteristics in an applied magnetic field. For
given value of the transport current $I^{^{\prime }}$ and changing of the
magnetic flux $\Phi _e$ the average voltage periodically oscillates with
amplitude $\delta V$ . For the detailed analysis of this behaviour in the fluctuation
free case see ref. \cite{oov2}.

\section{ Magnetic flux switching in the bistable state due to the noise and
transport current.}

We now use the numerical model to study the behaviour of the embraced 
magnetic flux in the ring.

Let us start with the fluctuation free case. For the time independent
control parameters $I$ and $\phi _e$ , we can introduce a potential:

\begin{equation}
U(\theta ,\phi ,\chi \mid I,\phi _e)=\frac{(\phi -\phi _e)^2}{2{\cal L}}%
-I\theta -\cos {}^2\frac \theta 2-\cos {}^2\frac \phi 2-2\cos \frac \theta
2\cos \frac \phi 2\cos \chi 
\end{equation}
such that the dynamical eqs.(6) take the form

\begin{equation}
\dot \theta =-\frac{\partial U}{\partial \theta },\dot \phi =-\frac{\partial
U}{\partial \phi },\dot \chi =-\frac 12\frac{\partial U}{\partial \chi } 
\end{equation}

The stable steady states of the system correspond to the minima of potential 
$U$ with respect to variables $\theta ,\phi ,\chi $ at given values of $I$
and $\phi _e$. The minimization of $U$ with respect to $\chi $ gives that
phase $\chi $ takes the value $0$ or $\pi $, depending on the equilibrium
values of $\theta $ and $\phi $ : $\cos \chi =sign(\cos \frac \theta 2\cos
\frac \phi 2)$. The stable configurations of the noise-free 4-terminal SQUID
where studied in ref. \cite{oov1} . It was shown that at the steady-state
domain in the plane of $(I,\phi _e)$ the region exists, inside which the
potential (10) has two minima (see Fig.3 in ref.\cite{oov2}). The structure
of this bistable state, the height and the width of the potential barrier in
the 3-dimensional phase space ($\theta ,\chi ,\phi $) can be regulated by
control parameters $I$ and $\phi _e$. For example, at $\phi _e=\pi $ we have
two equilibrium values of induced magnetic flux ,which are equally located
around $\pi $: $\phi _{1,2}(I)=\pi \pm \Delta \phi $. The current circulating in the
ring in this case equals $\pm j.$ The values of the potential $U$ at
these two states are equal, $U_1(I,\phi _e=\pi )=U_2(I,\phi _e=\pi )$, and
in which state the SQUID is located depends on the history of the system. It
is emphasised that in contrast to the case of the usual SQUID, the described
bistable state of the 4-terminal SQUID exists for each value of inductance $%
{\cal L}${\cal \ }even for ${\cal L}<1$.

The presence of noise, {\it i.e}. the three random forces $\delta v(t)$ in
the dynamical eqs.(6), produces small fluctuations of phases $\phi ,\theta
,\chi $ near the equilibrium values, as well as transitions between the two
states. Each such transition corresponds to the switching of the magnetic
flux $\phi $ between the two values $\phi _1$ and $\phi _2$. The switching
of magnetic flux must be accompanied by a switching of $\chi $ with a value $%
\pm \pi $ or by a $2\pi $ slippage of $\theta $. Note, that the phases $%
\theta $ and $\chi $ are independent parameters, having  different physical
meaning. In accordance with the eqs.(9) the dynamics of $\theta $ determines
the voltage in the transport channel, while the time dependence of $\chi $ means
the appearance of the voltage between the ring and the transport circuit. In
other words, the last one means the electrical charging of the ring. The
random event of one from the two possible processes, switching of $\theta $
or $\chi $ , is determined by the magnitudes of the corresponding potential
barriers $\Delta U_\theta $ and $\Delta U_\chi $. They depend on the value
of the transport current $I$. By analyzing the shape of the potential $%
U(\theta ,\phi ,\chi \mid I,\phi _e)$ surfaces can qualitatively be concluded (see
ref.\cite{bert}) that for small values of $I$ the transitions with switching of $\chi 
$ are more probable than slippage of $\theta $, but with the inreasing of
the current $I$ the situation will changes to the opposite.

To study the processes of switching in the bistable state, produced by the
thermal noise, we numerically solve the eqs.(6) and obtain the time
dependencies $\phi (t),\theta (t),\chi (t)$ . Figs.5 and 6 show some typical
traces of the phases $\phi ,\chi $ and $\theta $ as a function of time $t$
at an external flux value $\phi _e=\pi $ for particular values of the
noise parameter $\Gamma =0.2$ and of the selfinductance ${\cal L}=1.7.$

Fig.5 shows the behaviour for the transport current $I=0$. In this case we
observe the noisy behaviour of phases and the random switching of $\phi $
between two flux states. When $\phi $ switches, the phase $\chi $ jumps at
the same moment in time, with a value of $\pm \pi $. In the same time,
although the behaviour of $\theta (t)$ is noisy, no switching is observed in 
$\theta $.

In Fig.6 the transport current equals $I=0.25$. It displays the case , in
which the switching of $\phi $ between both flux states may be accompanied
by a jump of $\chi $ with a value of $\pm \pi $ or a $2\pi $ phase slippage
of $\theta $. In addition, in the numerical simulations the case is
observed, that $\chi $ and $\theta $ jump at the same time while $\phi $
remains constant.

The behaviour observed in Figs.5 and 6 corresponds to the above mentioned
dependence on the transport current of the potential barriers $\Delta
U_\theta $ and $\Delta U_\chi $. The switching of the magnetic flux can be
induced not only by the noise but also by the transport current $I$, if it
exceeds the critical value $I_c(\phi _e)$. Fig.7 shows the time dependencies
of the phases $\phi ,\chi ,\theta $ in the noise-free case ($\Gamma =0)$ for
the value of the current $I=1.05$. This value only very slightly
exceeds the critical current (see Fig.3) and we see that the flux $\phi $
adiabatically follows $\theta (t)$, except the moments of time when the
slippage of $\theta $ takes place and the corresponding  switch in the
flux occurs. In the same time, in this regime, the phase $\chi $ remains
constant.
\section{Summary}

The four terminal SQUID presents a system with three degrees of freedom,
the phase differences $\phi ,\theta ,\chi .$ In the steady state domain the
observable quantity is the magnetic flux embraced in the ring, $\Phi =\frac
\hbar {2e}\phi .$ In some region of the independent control parameters,
the transport current $I$ and the applied magnetic flux $\phi _e$, the system is in
the bistable state with two magnetic flux values. The operation of the four
terminal SQUID, {\it i.e} the choosing of one from two possible states, can be
achieved by variation of $I$ and $\phi _e$. At fixed values of the control
parameters, the magnetic flux switching is produced by the thermal noise or by
the transport current, which exceeds the critical value. 

We have computed the transitions between two flux states stimulated by the
thermal fluctuations.The switches of the flux are accompanied by the jumps
in $\theta $ or $\chi $ and the corresponding voltage impulses. The
existence of a bistable state in a four terminal SQUID was experimentally confirmed
in ref.\cite{bert}. by the direct observations of the magnetic flux behaviour, 
which is similar to the results obtained numerically in Section 4.

\newpage

FIGURE CAPTIONS \ \ \ \

Figure 1. \ \ The 4-terminal SQUID. The area of the region closed by the dashed lines 
(the Josephson 4-terminal junction) is of
the order of the coherence length squared. \ \

Figure 2.  \ \ A comparison of the numerically calculated noise-rounded $IV$ 
characteristics (points) for a single resistively shunted junction with the analytical
results from ref.\cite{amb} (solid lines). The dashed line is the noise-free 
characteristic. \ \

Figure 3. \ \ Current- voltage characteristics of the current-biased transport circuit
of a 4-terminal SQUID in the presence of noise for ${\Gamma}=0.2, {\cal L}=1.7$ and for two external flux
values ${\phi}^{e}=0$ (a) and ${\phi}^{e}={\pi}$ (b). The solid lines are the 
noise-free characteristics.\ \

Figure 4. \ \ The changing of the noise-rounded current-voltage 
characteristic of a 4-terminal SQUID in an applied magnetic field. ${\delta}V$ shows
the amplidude of the voltage oscillations as a function of 
the external flux ${\Phi}^{e}$ for an applied
transport current $I^{'}$. \ \

Figure 5. \ \ The three phase differences $\phi$, $\chi$ and $\theta$ 
versus time $t$ for ${\phi}^{e}={\pi}$, ${\cal L}=1.7$ in the presence of noise with 
${\Gamma}=0.2$. The transport current $I=0$. Transitions between the two flux 
states occur (jumps in ${\phi}$) which are 
accompanied by jumps in ${\chi}$ by ${\pm}{\pi}$. At zero transport current no switching
in ${\theta}$ occurs. \ \

Figure 6. \ \ This figure can be compared with Fig.5 but now the transport 
current is equal to $I=0.25$. Again ${\phi}^{e}={\pi}$. This is an intermediate situation where the
switching of ${\phi}$ is accompanied by jumps in ${\chi}$ by ${\pm}{\pi}$ 
or in ${\theta}$ by $2{\pi}$. At higher currents the ${\chi}$-fluctuation mechanism
will eventually be taken over by the ${\theta}$-fluctuation mechanism. \ \

Figure 7. \ \ The time dependencies of the phase differences $\phi$, $\chi$
and $\theta$ in the noise-free case (${\Gamma}=0$) for the value of the
current $I=1.05$, which slightly exceeds the critical current and for ${\phi}^{e}={\pi}$.
At the moments 
of time when the $2{\pi}$ phase slippage of ${\theta}$ takes place the switch
in the flux ${\phi}$ occurs. The phase ${\chi}$ in this regime remains constant. \ \
\end{document}